\documentstyle[12pt]{article}
\begin{document}
\title{THE UNIFICATION OF ELECTROMAGNETISM AND GRAVITATION IN THE CONTEXT
OF QUANTIZED FRACTAL SPACE TIME}
\author{B.G. Sidharth$^*$\\
Centre for Applicable Mathematics \& Computer Sciences\\
B.M. Birla Science Centre, Adarsh Nagar, Hyderabad - 500 063 (India)}
\date{}
\maketitle
\footnotetext{$^*$Email:birlasc@hd1.vsnl.net.in; birlard@ap.nic.in}
\begin{abstract}
The attempts to unify electromagnetism and gravitation have included the
formulations of Herman Weyl and the Kaluza Klein theory with the fifth
dimension. More recently there have been fruitful attempts in the domain
of Quantum Superstrings and the author's formulation in terms of Quantum
Mechanical Kerr-Newman Black Holes. Though all these appear to be widely
divergent approaches, they are shown to have a unified underpinning in the
context of quantized fractal space time.
\end{abstract}
\section{Introduction}
The problem of the unification of gravitation and electromagnetism has a long
and as yet unconcluded history. The very early attempt of Weyl did not find favour
\cite{r1} as he had incorporated electromagnetism into the field equation, as
it were from outside\cite{r2}. The ingenous suggestion of Kaluza that a
fifth, and somehow suppressed, dimension be introduced \cite{r3,r1}
also did not find favour, though the idea has resurfaced in the theory of Quantum
Superstrings\cite{r4}. Another approach has been that of Quantum Mechanical
Black Holes which recovers the Kerr-Newman metric in the context of quantized
fractal space time (QFST)\cite{r5,r6,r7}. While there is formal resemblance
to the Weyl formulation on the one hand, there is an imaginary shift of
coordinates, quite meaningful in the above context, on the other. In the
derivation of the Kerr Newman metric too\cite{r8}, there is a mysterious imaginary
shift which however is not explicable in classical theory. Finally both in
Quantum Superstrings and the (QFST) or quantized fractal space time picture, the underlying
geometry is non-commutative \cite{r9,r10,r4}.\\
So there are in these scenarios several at first sight disparate and even
inexplicable strands: the Kaluza-Klein curled up extra dimension, Weyl's
electromagnetism, Quantum Mechanical Black Holes, Quantum Superstrings, the
Kerr-Newman metric, non-commutative geometry.... We will now show that infact
all these characteristics have a unified underpinning in the context of
quantified fractal space time (QFST).
\section{The Kaluza-Klein and Weyl Formulations}
Our starting point is the fact that the fractal dimension of a Brownian quantum
path is 2, as pointed out by Abbott and Wise, Nottale and others\cite{r11,r12}.
This was further analysed by the author and it was explained
that this is symptomatic of quantized fractal space time and it was shown that
infact the coordinate $x$ becomes $x+\imath ct$\cite{r13}, reminiscent of El Naschie's
complex time\cite{r14} and the Hawking-Hartle static time\cite{r15}. The
complex coordinates or equivalently non-Hermitian position operators are
symptomatic of the unphysical zitterbewegung which is eliminated after an averaging
over the Compton scale. In this picture the fluctuational creation of particles is
taken into account in a consistent cosmological scheme\cite{r16}.\\
It is well known that the generalization of complex $x$ coordinate to three
dimensions leads to quarternions\cite{r17}, and the Pauli spin metrics.\\
We next come to a model of an electron as a Quantum Mechanical Black Hole and
widely discussed by the author (Cf. example\cite{r5,r6}). In this model
an electron is a spinning shell (reminiscent of Dirac's model\cite{r18})
of radius equalling the Compton wavelength, and equavalently a Kerr-Newman
Black Hole. Within this Compton scale Black Hole we encounter the unphysical
zitterbewegung region of complex (or non-Hermitian) coordinates already alluded to.\\
Infact electromagnetism was deduced\cite{r5,r6,r7} in two
ways. The first was by considering an imaginary shift,
\begin{equation}
x^\mu \to x^\mu + \imath a^\mu , (a^\mu \sim \mbox{Compton scale})\label{e1}
\end{equation}
in a Quantum Mechanical context. This lead to
\begin{equation}
\imath \hbar \frac{\partial}{\partial x^\mu} \to \imath \hbar \frac{\partial}
{\partial x^\mu} + \frac{\hbar}{a^\mu}\label{e2}
\end{equation}
and the second term on the right side of (\ref{e2}) was shown to be the
electromagnetic vector potential $A^\mu$,
\begin{equation}
A^\mu = \hbar/a^\mu\label{e3}
\end{equation}
The second was by taking into account the fact that at the Compton scale, it
is the so called negative energy two spinors $\chi$ of the Dirac bispinor that
dominate where,
$$\chi \to -\chi$$
under reflections. This lead to the tensor density property,
\begin{equation}
\frac{\partial}{\partial x^\mu} to \frac{\partial}{\partial x^\mu} - \Gamma_\nu^{\mu \nu}\label{e4}
\end{equation}
the second term on the right side of (\ref{e4}) being identified with $A^\mu$,
\begin{equation}
A^\mu = \hbar \Gamma_\nu^{\mu \nu}\label{e5}
\end{equation}
It was pointed out that (\ref{e5}) is formally and mathematically identical to
Weyl's original formulation\cite{r1}, except that here it arises due to the
purely Quantum Mechanical spinorial behaviour whereas Weyl had put it by hand.\\
Another early scheme for the unification of gravitation and electromagnetism
as referred to earlier was that put forward by Kaluza and Klein\cite{r3} in which
an extra dimension was introduced and taken to be curled up. This idea has
resurfaced in recent years in String Theory.\\
We will first show that the characterization of $A^\mu$ in (\ref{e3}) is
identical to a Kaluza Klein formulation. Then we will show that equations (\ref{e4})
and (\ref{e5}) really denote the fact that the geometry around an electron is
non-integrable. Finally we will show that infact both (\ref{e2}) or (\ref{e3})
and (\ref{e4}) or (\ref{e5}) are the same formulations.\\
We first observe that the transformation (\ref{e5}) can be written as,
\begin{equation}
x^\imath \to x^\imath + \alpha_{\imath 5} x^5\label{e6}
\end{equation}
where $\alpha_{\imath 5}$ in (\ref{e6}) will represent a small shift from the
Minkowski metric $g_{\imath j}$, and $\imath , j=1,2,3,4,5,x^5$ being a fifth
coordinate introduced for purely mathematical conversion.\\
Owing to (\ref{e6}), we will have,
\begin{equation}
g_{\imath j} dx^\imath dx^j \to g_{\imath j} dx^\imath dx^j + (g_{\imath j}
\alpha_{j^5})dx^\imath dx^5\label{e7}
\end{equation}
In Kaluza's formulation,
\begin{equation}
A_\mu \propto g_{\mu 5}\label{e8}
\end{equation}
Comparison of (\ref{e6}) and (\ref{e7}) with (\ref{e1}) and (\ref{e3}) shows
that indeed this is the case. That is, the formulation given in (\ref{e1})
and (\ref{e2}) could be thought of as introducing a fifth curled up
dimension, as in the Kaluza-Klein theory.\\
To see why the Quantum Mechanical formulation (\ref{e4}) and (\ref{e5})
corresponds to Weyl's theory, we start with the effect of an infinitesimal
parallel displacement of a vector [Bergmann].
\begin{equation}
\delta a^\sigma = -\Gamma^\sigma_{\mu \nu} a^\mu d x^{\nu}\label{e9}
\end{equation}
As is well known, (\ref{e9}) represents the extra effect in displacements,
due to the curvature of space - in a flat space, the right side would vanish.
Considering partial derivatives with respect to the $\mu^{th}$
coordinate, this would mean that, due to (\ref{e9})
\begin{equation}
\frac{\partial a^\sigma}{\partial x^\mu} \to \frac{\partial a^\sigma}{\partial x^\mu}
- \Gamma^\sigma_{\mu \nu} a^\nu\label{e10}
\end{equation}
The second term on the right side of (\ref{e10}) can be written as:
$$-\Gamma^\lambda_{\mu \nu} g^\nu_\lambda a^\sigma = -\Gamma^\nu_{\mu \nu} a^\sigma$$
where we have utilized the property that in the above formulation (Cf.refs.[5,6,7]),
$$g_{\mu \nu} = \eta_{\mu \nu} + h_{\mu \nu},$$
$\eta_{\mu \nu}$ being the Minkowski metric and $h_{\mu \nu}$ a small
correction whose square is neglected.\\
That in, (\ref{e10}) becomes,
\begin{equation}
\frac{\partial}{\partial x^\mu} \to \frac{\partial}{\partial x^\mu} -
\Gamma^\nu_{\mu \nu}\label{e11}
\end{equation}
The relation (\ref{e11}) is the same as the relation (\ref{e4}).\\
We will next show the correspondence between (\ref{e11}) or (\ref{e5})
or (\ref{e4}) and (\ref{e3}) or (\ref{e2}). To see this simply we note that the
geodesic equation is,
\begin{equation}
\dot u^\mu \equiv \frac{du^\mu}{ds} = \Gamma^\mu_{\nu \sigma} u^\nu u^\sigma\label{e12}
\end{equation}
We also use the fact that in the Quantum Mechanical Black Hole model referred
to, we have [5]
$$u^\mu = c \quad \mbox{for}\quad \mu = 1, 2 \quad \mbox{and}3,$$
while,
$$|\dot u^\mu | = |u^\mu | \frac{mc^2}{\hbar}$$
So, from (\ref{e12}) we get,
$$\Gamma^\mu_{\nu \mu} = \frac{1}{a^\nu} , |a^\nu | = \frac{\hbar}{mc}$$
This establishes the required identity.
\section{Quantized Fractal Space time and Quantum Superstrings}
It was shown\cite{r9} that the quantized fractal space time referred to really
leads to a non-commutative geometry, not surprisingly:
\begin{equation}
[x,y] = 0(l^2),[x,p_x] = \imath \hbar [1 + l^2], [t,E] = \imath \hbar
[1+\tau^2],\cdots\label{e13}
\end{equation}
It was shown earlier\cite{r9} that these relations directly lead to the Dirac equation:
Quantized fractal space time or the above relation (\ref{e13}) are
the underpinning for Quantum Mechanical spin or the Quantum Mechanical Black
Hole, that is ultimately equations like (\ref{e2}) or (\ref{e3}) and
(\ref{e4}) or (\ref{e5}).\\
It is also true that both the Kaluza Klein formulation and the non commutative
geometry (\ref{e13}) hold in the theory of Quantum Superstrings.\\
Infact we get from here a clue to the mysterious six extra curled up dimensions
of Quantum Superstring Theory. For this we observe that (\ref{e13}) gives
an additional contribution to the Heisenberg Uncertainity Principle and we
can easily deduce
$$\Delta p \dot \Delta x \sim \hbar l^2$$
Remembering that at this Compton scale
$$\Delta p \sim mc$$
It follows that
\begin{equation}
\Delta x \sim l^3\label{e14}
\end{equation}
as $l \sim 10^{-11}cms$ for the electron we recover from (\ref{e14}) the
Planck Scale, as well as a rationale for the peculiar fact that the Planck
Scale is the cube of the electron Compton scale.\\
More importantly, what (\ref{e14}) shows is, that at this level, the single
dimension along the $x$ axis shows up as being three dimensional. That is there
are two extra dimensions, in the unphysical region below the Compton scale.
As this is true for the $y$ and $z$ coordinates also, there are a total of six
curled up or unphysical or inaccessible dimensions in the context of the
preceding section.
\section{Discussion and Conclusion}
If we start with equations (\ref{e1}) to (\ref{e3}) which were related to QFST
(Quantized Fractal space time) and the non-commutative relation (\ref{e13})
we obtain a unification of electromagnetism and gravitation. On the other hand
if we consider the spinorial behaviour of the Dirac wave function, we get
(\ref{e4}) or (\ref{e5}). The former has been seen to be the same as the
Kaluza formulation while the latter is formally similar to the Weyl formulation -
but in this case (\ref{e5}) is not put in by hand. Rather it is a Quantum
Mechanical consequence. We have thus shown that these two approaches are the
same. The extra dimensions are thus seen to be confined to the unphysical
Compton scale - classically speaking they are curled up or inaccessible.\\
In a sense this is not surprising. The bridge between the two approaches was
the Kerr-Newman metric which uses, though without a clear physical meaning in
classical theory, the transformation (\ref{e1}). The reason why an imaginary
shift is associated with spin is to be found in the Quantum Mechanical
zitterbewegung and the consequent QFST.\\
Wheeler remarked \cite{r19}, "the most evident shortcoming of the geometrodynamic
model as it stands is this, that it fails to supply any completely natural
place for spin $1/2$ in general and for the neutrino, in particular", while
"it is impossible to accept any description of elementary particles that does
not have a place for spin half." Infact the bridge between the two is the
transformation (\ref{e1}). It introduces spin half into general relativity
and curvature to the electron theory, via the equation (\ref{e5}) or (\ref{e10}).\\
In this context it is interesting to note that El Naschie has given the fractal
formulation of gravitation\cite{r20}.\\
Thus apparently disparate concepts like the Kaluza Klein and Weyl formulations, Quantum Mechanical Black
Holes, quantized fractal space time and QSS are seen to have a harmonius
overlap, in the context of QFST with its roots in the fluctuational creation
of particles\cite{r21}.

\end{document}